\documentclass[11pt,twoside]{article}

\usepackage{baaa}
\usepackage{graphicx}
\usepackage{subfigure}
\usepackage{amssymb}
\usepackage[spanish,activeacute,english]{babel}
\usepackage[latin1]{inputenc}
\usepackage{verbatim}
\usepackage{amsmath}
\usepackage{amsfonts}
\usepackage{amssymb}
\usepackage[colorlinks=true,dvips]{hyperref}
\usepackage{natbib}

\begin{document}
%%%%%%%%%%%%%%%%%%%%%%%%%%%%%%%%%%%%%%%%%%%%%%%%%%%%%%%%%%%%%%%%%%%%%%%%%%
%%%% SELECCIONE EL IDIOMA EN QUE SE ESCRIBE EL ARTCULO:              %%%%
%\myselectspanish
\myselectenglish
%%%%%%%%%%%%%%%%%%%%%%%%%%%%%%%%%%%%%%%%%%%%%%%%%%%%%%%%%%%%%%%%%%%%%%%%%%
\vskip 1.0cm
\markboth{J.C.N. de Araujo}
{Ondas gravitacionales y objetos compactos}

\pagestyle{myheadings}
%%%% DESCOMENTE LA LINEA QUE DESCRIBE EL CARACTER DE SU TRABAJO       %%%%
\vspace*{0.5cm}

\noindent TRABAJO INVITADO
%\noindent PRESENTACIN ORAL
%\noindent PRESENTACIN MURAL
%\noindent RESUMEN

\vskip 0.3cm
\title{Ondas gravitacionales y objetos compactos}

%\title{ Template paper for publication in the Bulletin of the
%Argentinian Astronomical Association with instructions for the use of
%\LaTeX{}}

\author{J.C.N. de Araujo$^{1}$}

\affil{(1) Divis\~ao de Astrof\'{\i}sica - Instituto Nacional de Pesquisas Espaciais(Brazil)}

\begin{abstract}
It is presented a brief review on gravitational waves (GWs). It is shown how the wave equation is obtained from Einstein's equations and how many and how are the polarization modes of these waves. It is discussed the reasons why GWs sources should be of astrophysical or cosmological origin. Thus, it is discussed what would be the most likely sources of GWs to be detected by the detectors of GWs currently in operation and those that should be operational in the future, emphasizing in particular the sources involving compact objects. The compact objects such as neutron stars, black holes and binary systems involving compact stars can be important sources of GWs. Last but not least, it is discussed the GWs astrophysics that is already possible to do, in particular involving the compact objects.
\end{abstract}

\begin{resumen}
  Es presentada una breve revisi\'{o}n sobre las ondas gravitacionales (OGs).
Mostramos como es obtenida la ecuaci\'{o}n de onda a partir de las ecuaciones
de Einstein adem\'{a}s de cu\'{a}ntos y c\'{o}mo son los modos de polarizaci\'{o}n de esas
ondas. Son discutidas las razones por las cu\'{a}les las fuentes de OGs deben
ser de origen astrof\'{i}sico o cosmol\'{o}gico. As\'{i}, pasamos a discutir cu\'{a}les
ser\'{i}an las fuentes de OGs m\'{a}s probables a ser detectadas
por los detectores de OGs que actualmente est\'{a}n en operaci\'{o}n y aquellos
que deben entrar en operaci\'{o}n en el futuro, haciendo \'{e}nfasis en particular
en las fuentes que envuelven objetos compactos. Los objetos compactos,
tales como las estrellas de neutrones, agujeros negros, as\'{i} como los
sistemas binarios que envuelven estrellas compactas pueden ser fuentes importantes
de OGs. Por \'{u}ltimo, discutimos el hecho, no menos
importante, de que ya es posible hacer astrof\'{i}sica de OGs, en particular con los objetos compactos.
\end{resumen}

\section{Introduction}
\label{S_intro}

According to General Relativity there exists the so called GWs
\cite[see, e.g.,][]{1987thyg.book..330T}. These waves are ripples in
spacetime curvature that propagate with the speed of light.

As is well known, General Relativity is an extremely well succeeded theory.
However, to be completely well succeeded, the GWs must be detected and their
characteristics must be, necessarily, in accordance with the previsions of this theory.

Since the 1960s there is a great effort to detect GWs
\cite[see, e.g.,][]{2013PrPNP..68....1R}.  However, the sensitivities of the various experiments along these five decades have not yet allowed a direct detection. This is so also because the amplitudes of the GWs are extremely small.

On the other hand, the observed inspiral rate of the binary
pulsar PSR1913+16 (Hulse-Taylor binary)\footnote{There is only a few known neutron star-neutron star
(NS-NS) binaries, among them the double pulsar system J0737-3039 discovered
in 2003 \cite{2004Sci...303.1153L}.} and its excellent agreement with the prediction
given by General Relativity provide a strong indirect evidence of
the existence of GWs \citep{1975ApJ...195L..51H}.

However, it is quite likely that during the current decade it will be
finally possible to detect GWs directly, because the sensitivities of experiments
such as the advanced configurations of LIGO and VIRGO are about to
reach a level that will finally allow detections.

%Therefore, it is possible to say that we are about to
%enter into the era of the Gravitational Wave Astronomy.

Concerning the sources of GWs, they will be of astronomical
and cosmological nature, since it is easy to be convinced,
as we are going to see later on, that it is not possible
to generate in laboratory GWs at detectable level.
Also, the astronomical sources of GWs, in particular
the most probable to be detected in the near future,
involve compact objects.

It is worth bearing in mind that it is impossible to cover in the
present contribution all aspects involving GWs.
It will be given a general vision without entering
in details of the many aspects involving such an issue.

Without being pretentious, my intention here, as in my talk at AAA 2012,
is to motivate other researchers to be interested in this extraordinary
area of research.

This contribution is organized as follows: in Section \ref{GW}, it is
shown how to obtain the GW equation, its polarization modes, etc; in
Section \ref{SOURWG} it is considered the main sources of GWs; in Section \ref{DETGW} it is briefly considered the detection of GWs; Section \ref{GWASTRO} deals with the GW astronomy that is already
possible to do;  and, finally, in Section \ref{Final} it is presented the final remarks.

\section{Gravitational wave equation, polarization modes, etc}
\label{GW}

The detailed derivation of the GW equation in the contest of General Relativity
can be found in many texbooks, we refer the reader to the well known book by
\cite{MTW73}. We here consider only the main steps of this derivation.

The GW equation comes from the linearization of the
General Relativity equations. Therefore, the starting point reads

\begin{equation}
\label{gre}
R_{\alpha\beta} - \dfrac{1}{2}g_{\alpha\beta}R= \dfrac{8\pi G}{c^{4}}T_{\alpha\beta},
\end{equation}

\noindent where $R_{\alpha\beta}$ is the Ricci tensor, $R$ is the Ricci scalar, and $T_{\alpha\beta}$ is the energy momentum tensor. The other quantities refer to well known physical constants. Notice that the Greek (Latin) indexes run from 0 to 3 (1 to 3).

Now, consider the weak field approximation, where the metric $g_{\alpha\beta}$ can be written as

\begin{equation}
\label{metr}
g_{\alpha\beta} = \eta_{\alpha\beta} + h_{\alpha\beta},
\end{equation}

\noindent where $\eta_{\alpha\beta}$ is the Minkowsky metric (flat spacetime) and $h_{\alpha\beta}$ is a small perturbation.

Substituting equation (\ref{metr}) into equation (\ref{gre}), defining
$\bar{h}_{\alpha\beta} = h_{\alpha\beta} - \dfrac{1}{2} h \eta_{\alpha\beta}$,
and considering the gauge $\bar{h}^{\alpha\beta}_{\beta}=0$ (Lorentz gauge) one
obtains the wave equation, namely

\begin{equation}
\Box\bar{h}_{\alpha\beta}= -16\pi \dfrac{G}{c^{4}} T_{\alpha\beta}.
\end{equation}

To study the properties of the GWs one can look at the plane waves, namely

\begin{equation}
\bar{h}_{\alpha\beta}= A_{\alpha\beta}\exp(2\pi ik_{\mu}x^{\mu}),
\end{equation}

\par\noindent where the amplitudes ($A_{\alpha\beta}$) and the wave vector
($k_{\mu}$) are constants. The field equations imply that the wave vector is null
($k_{\mu}k^{\mu}$=0), therefore the GWs propagate at speed of light. The gauge condition
imply that $A_{\alpha\beta}k^{\beta}=0$.

Further gauge conditions can be applied, in particular the so called transverse and traceless gauge, or TT gauge, namely

\begin{equation}
h^{0\alpha}= 0,\qquad h^i_i=0, \qquad \partial^ih_{ij}=0
\end{equation}

\par\noindent \cite[see, e.g.,][for a detailed discussion]{Maggiore_Book}.

Notice that from the above discussion one can also conclude that
$A^{ij}k_j=0$, i.e., the GWs are transverse, as with the electromagnetic
waves. Also, counting the independent components of $A^{\alpha\beta}$ after
applying the gauge conditions one concludes that the GWs have two polarizations.

It is worth noting that it can be shown that TT gauge represents a coordinate
system that is comoving with freely falling particles. As a result, there
is no effect of a GW in TT gauge on a particle at rest, i.e., the particle
remains at rest.

On the other hand, a GW affects, for example, the separation of two freely falling particles, i.e., a GW generates a tidal effect.

Now, considering a plane wave propagating in the z-direction, one can conclude
that the only non-null components of $A^{\alpha\beta}$ are $A^{xx}$, $A^{xy}=A^{yx}$
and $A^{yy}$; with $A^{xx}=-A^{yy}$.

Then, the two polarizations modes can be explicitly shown, writing the
associated metrics, namely

\begin{equation}
ds^2 = -dt^2 + (1+h_+)dx^2 + (1-h_+)dy^2 + dz^2
\end{equation}

\par\noindent where $ h_+ \equiv  A^{xx} \exp[ik(z-t)]$ for a GW with $A^{xy}=0$; and

\begin{equation}
ds^2 = -dt^2 + dx^2+ 2h_\times dxdy + dy^2 + dz^2
\end{equation}

\par\noindent where $h_\times \equiv A^{xy}\exp[ik(z-t)]$ for a GW with $A^{xx}=0$.
Note that this last metric can be obtained from the previous one by performing a rotation of 45 degrees.

\begin{figure}[!ht]
  \centering
  \includegraphics[width=0.45\textwidth, angle=-90]{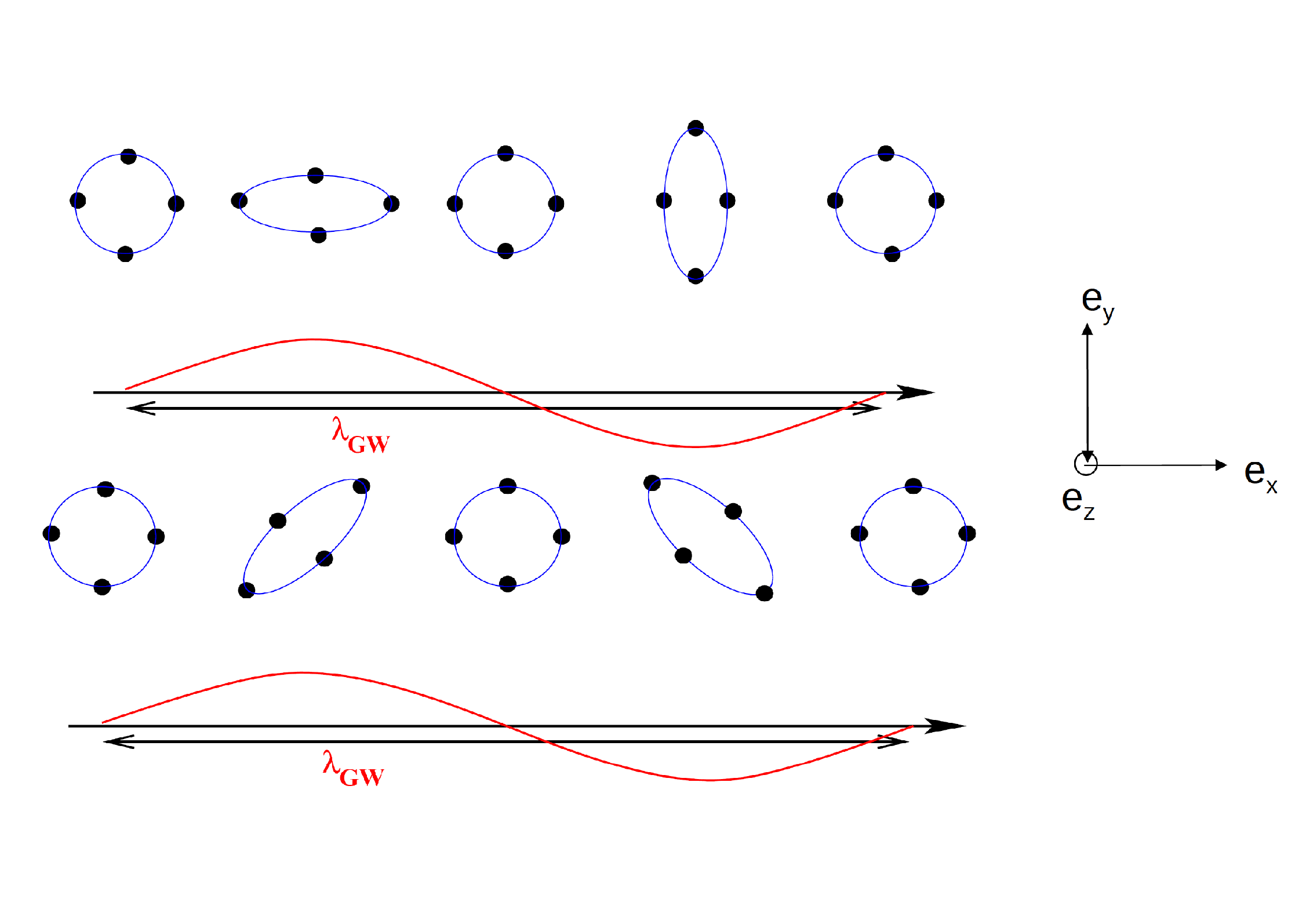}

  \caption{The deformation of a ring of test masses due to the $+$ and $\times$ polarizations.}
  \label{anel}
\end{figure}

In figure \ref{anel} it is shown the deformation of a ring of test masses, located in xy-plane, caused by a GW that propagates along the z-direction.

Notice that, when a GW of amplitude $h$ ($=\sqrt{h_{+}^{2}+h_{\times}^{2}}$) hits perpendicularly a detector of length scale L, it generates a length change of $\delta L/L \sim h/2$.

In order to know the nature of the sources that generate GWs
it is appropriate to decompose them into their multipolar contributions.

Performing such an expansion, one obtains no monopolar or dipole contributions. Energy conservation prohibits monopole gravitational radiation. Conservation of linear and angular momenta prohibit
dipolar gravitational radiation. The minimum contribution is quadrupolar, namely

\begin{equation}\label{hquad}
  h^{TT}_{ij} = \frac{2}{r} \frac{G}{c^{4}} \ddot{Q}^{TT}_{ij}(t-r/c),
\end{equation}

\par\noindent where the quadupole momentum reads,

\begin{equation}\label{quadm}
  \ddot{Q}^{TT}_{ij} = \int\rho(x^{i}x^{j}-\frac{1}{3}\delta^{ij}r^{2})d^{3}x,
\end{equation}

\par\noindent where $\rho$ is the rest mass density. Notice that
the above equation holds for slow internal motion and weak field.

The $G/c^4$ in equation \ref{hquad} is extremely small, therefore to
have detectable effects the second time derivative of the quadrupole
momentum must be enormous.

In the next section it is considered the possible sources of GWs, and if it is possible to generate in
laboratory  detectable (manmade) sources.

\section{Sources of Gravitational waves}
\label{SOURWG}

\begin{figure}[!ht]
  \centering
    \includegraphics[width=0.45\textwidth]{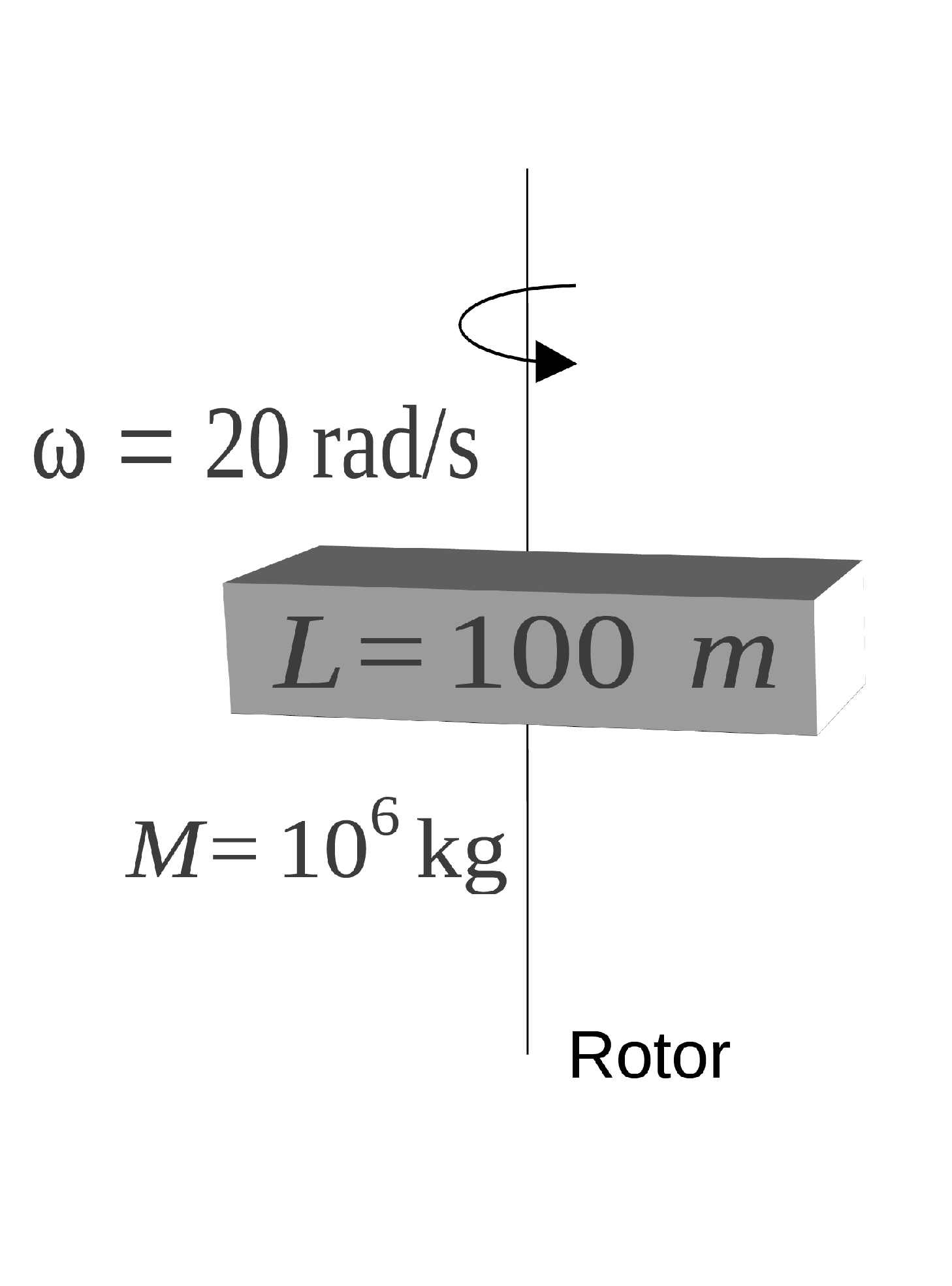}
  \caption{A ``laboratory experiment" to produce GWs.}
  \label{rotor}
\end{figure}

One could ask if it is possible to produce GWs in laboratory.
To answer this question one could consider a ``laboratory experiment"
where one uses, for example, a bar with M = $10^6$ kg  and  L = 100 m,
rotating, say, at $\omega =20$ rad/s (see figure \ref{rotor}).
It can be shown, using for example equation \ref{hquad}, that the amplitude
of the GW for this rotor is given by

\begin{equation}\label{hrotor}
h  \sim 10^{-45} \frac{ML^{2}\omega^2}{r},
\end{equation}

\par\noindent where r is the distance source-detector.
To detect the GW generated by this experiment, however,
one should be at the wave zone, i.e., at a
distance $r > \lambda_{GW}$ ($\simeq 50,000$ km, in the present case).
This implies that the amplitude h reads

\begin{equation}
h < 10^{-40};
\end{equation}

\par\noindent which is absolutely impossible to be detected.

Since it is not possible to generate GWs in laboratory at
a detectable level, the sources of GWs should be of astronomical
and cosmological nature. Even considering astronomical and
cosmological sources, the GWs that one is hoping to detect on
Earth have extremely small amplitudes.

Using the same example given by \cite{1994figw.book.....S},
consider a binary system formed by a pair of 1.4 $M_{\odot}$
NSs in a circular orbit of 20 km, and located at the Virgo
galactic cluster, which is $\sim$ 20 Mpc away. This system emit GWs
at 800 Hz with amplitude  $h \sim 10^{-21}$. This is the typical value
of the amplitudes of GWs generate by astrophysical sources. This
figure also explains why GWs have not yet been detected.

Besides, not only the amplitudes are small, but the number of sources
available for detection in the Galaxy, for example, is small.
Thus, it is necessary the highest possible sensitivity in order to put
other galaxies in to play and, as a result, a greater number of sources.

An useful equation that can obtained from equation (\ref{hquad}),
the quadrupole formula, reads

\begin{equation}\label{hc}
  h \sim \frac{G^{2}}{c^{4}} \Big(\frac{M}{r}\Big)\Big(\frac{M}{R}\Big)
\end{equation}

\par\noindent where $R$ is a characteristic dimension of the source.
This equation shows that M/R (the compactness) needs to be large for
the source be a promising source of GWs. That is why the most promising
sources of GWs involve compact objects, namely, NSs and black holes (BHs). It is worth
mentioning that supermassive black holes (SMBHs) are compact objects too.

In general, the sources are usually classified according to their typical pattern of emitted GWs. There are basically four types of sources: the periodic sources such as binary systems (with at least one compact object) at the early stages of their existence, and rotating NSs. In this cases the emitted signals are characterized by a regular behavior, with frequencies and amplitudes being constant over a long timescale; the second type are the burst sources, characterized by strong emissions at short time intervals, where the main sources are the merging of compact objects (NS-NS, NS-BH and BH-BH binaries), triggering of supernovae, and oscillating modes of NSs and BHs; a third type is called ``chirp", which is an intermediate case between the periodic and the burst signals and occur at the final stages of existence of compact binary systems, where the frequencies and amplitudes evolve from a periodic regime and undergo an exponential and very fast increase with the process ending in a burst; the fourth type refers to stochastic backgrounds, which are characterized by spectra spanning a wide range of frequencies. These spectra could have been generated by cosmological processes in the very early Universe, such as phase transitions and spatial inhomogeneities and also by a superposition of several individual sources, that may be of periodic, burst or chirp nature.

In general, the frequency bands where are expected to find GWs
span from $10^{-18}$Hz up to $10^{4}$Hz, given the above  (anticipated) sources. Serendipitous (unanticipated) sources could also well exit, generating GWs with frequencies inside or outside this frequency band.

 The~reader~should~look~at~the~webpage
 \href{http://www.ligo.org/science/GW-Sources.php}{http://www.ligo.org/science/GW-Sources.php}
 where interesting materials  about GWs can be found.

It is worth noting that GWs carry information about the dynamics of the sources.  Also, since GWs are weakly interacting, they
could well traverse the universe without being significantly scattered or absorbed.

\section{Detecting gravitational waves: present and future}
\label{DETGW}

The first experimental researches of GWs took place in the 1960s
with Joseph Weber \cite[see, e.g.,][and references therein]{1987thyg.book..330T}.
Since then, many other projects have been proposed
\cite[see, e.g.,][and references therein]{2011RMxAC..40..299A}.

Nowadays the search for GWs involves different techniques and experiments, namely,
resonant mass detectors (bars - such as in the pioneer experiment by
\cite{1960PhRv..117..306W}-, and spheres), interferometers, radio telescopes,
and cosmic background radiation experiments.

Resonant detectors (bars and spheres) absorb energy from the GW
and mechanically resonate. With electromechanical transducers
the mechanical signals are converted into electrical ones.
In general, this kind of detector is sensitive to GWs in a narrow band
of frequencies, i.e., at most $\sim 10\%$ of the resonant frequency
(1kHz and 3kHz for the bars and spheres in operation or under development,
respectively).

AURIGA and NAUTILIUS are examples of resonant bar detectors still in
operation. We refer the reader to their webpages
\href{http://www.auriga.lnl.infn.it/}{http://www.auriga.lnl.infn.it/}  and
\href{http://www.roma1.infn.it/rog/nautilus/}{http://www.roma1.infn.it/rog/nautilus/}, respectively)
for further details.

The schematics of a GW bar detector can be found for example in
\href{http://www.auriga.lnl.infn.it/auriga/detector/overview.html}{http://www.auriga.lnl.infn.it/auriga/detector/overview.html}.

Concerning~spherical~detectors,~SCHENBERG
(\href{http://www.das.inpe.br/graviton/}{http://www.das.inpe.br/graviton/})
and MiniGRAIL (\href{http://www.minigrail.nl/}{http://www.minigrail.nl/}) are
still under development.

A GW antenna with a spherical shape maximizes the GW absorption
and is omnidirectional (i.e., the same sensitivity in any direction).
They are still narrow band, but can be, in principle, more
sensitive than resonant bars \cite[see, e.g.,][]{2011RMxAC..40..299A}.

The interferometers are a kind of a giant Michelson-Morley experiment.
Making use of laser interferometry, such detectors measure differences
between the distances of test bodies caused by the passage of GWs.

The reader should look at the webpage
\href{http://www.ligo.org/science/GW-Overview/images/IFO.jpg}{http://www.ligo.org/science/GW-Overview/images/IFO.jpg},
where a diagram of a basic interferometer design can be found.

Examples of ground-based detectors are the following:

\par\noindent LIGO (\href{http://www.ligo.org/}{http://www.ligo.org/}) % LIGO

\par\noindent VIRGO (\href{http://www.virgo.infn.it}{http://www.virgo.infn.it}) % VIRGO

\par\noindent GEO600 (\href{http://www.geo600.org/}{http://www.geo600.org/}) % GEO600

\par\noindent TAMA300 (\href{http://tamago.mtk.nao.ac.jp/spacetime/}{http://tamago.mtk.nao.ac.jp/spacetime/})  % TAMA300

We refer the reader to the above indicated webpages for details.
The frequency band of these detectors ranges from $\sim 10^{2} - 10^{4}$ Hz.

Nowadays (2013), the most sensitive is LIGO, where the strain spectral density
is of $\simeq 2\times 10^{-23} Hz^{-1/2}$ for $\simeq 200$ Hz
(see the LIGO webpage for further details).

Notice that the interferometers are broad band detectors and can be more sensitivity
than resonant mass detectors.

In the future, a decade or so, there will be other ground-based detectors around, namely:

\par\noindent KAGRA (\href{http://gw.icrr.u-tokyo.ac.jp/lcgt/}{http://gw.icrr.u-tokyo.ac.jp/lcgt/}, already in construction) % LCGT or KAGRA

\par\noindent INDIGO (\href{http://www.gw-indigo.org}{http://www.gw-indigo.org})  % INDIGO

\par\noindent AIGO (\href{http://www.aigo.org.au}{http://www.aigo.org.au}) % AIGO

\par\noindent Einstein Telescope (ET, \href{http://www.et-gw.eu/}{http://www.et-gw.eu/}) % ET

Moreover, there will be the advanced configurations of LIGO and VIRGO.
The sensitivity of the advanced LIGO (aLIGO) will be ten times better than the initial
LIGO. This implies in an observable volume of the universe a thousand larger.

The main sources of GWs for the resonant detectors and the ground-based interferometers
are the compact objects.

For frequencies below 1 Hz or so, it  is not possible
to have a competitive sensitivity because of gradients
of the gravitational field. Therefore, the
detectors (antennas) must be necessarily space-based.

The main example of space-base antenna is LISA, which was until recently a
NSA/ESA project (see \href{http://lisa.nasa.gov/}{http://lisa.nasa.gov/}
and \href{http://sci.esa.int/lisa}{http://sci.esa.int/lisa}, for details).
Now ESA has its own project, the eLISA or NGO (\href{http://www.elisa-ngo.org/}{http://www.elisa-ngo.org/}),
which is less sensitive than the former LISA for $\sim 0.01$ Hz, but still scientifically competitive.

Other projects of space-based antennas are

\par\noindent DECIGO (\href{http://tamago.mtk.nao.ac.jp/decigo}{http://tamago.mtk.nao.ac.jp/decigo}) % DECIGO

\par\noindent BBO (\href{http://trs-new.jpl.nasa.gov/dspace/bitstream/2014/37836/1/05-2157.pdf}{http://trs-new.jpl.nasa.gov/dspace/bitstream/2014/37836/1/05-2157.pdf})
% BBO

The main sources of GWs for the space-based interferometers
are the supermassive black holes and stochastic waves.

Concerning the sensitivities of the space-based antennas, the eLIGO strain spectral density
for example is of $\sim 10^{-23} Hz^{-1/2}$ at $\simeq 0.01$ Hz (see the eLIGO webpage for further
details).

Pulsar Timing Arrays (PTAs) are a new way of searching GWs \cite[see, e.g.,][]{2011RMxAC..40..299A}.
The basic idea is to use extremely regular Pulsars
as a detector. Long wavelength GWs cause discrepancies
in the Pulsar pulses and this effect is used as a detector
(see, e.g., the webpage of the Parkes
Pulsar Timing Array - PPTA, the pioneer group,
\href{http://www.atnf.csiro.au/research/pulsar/ppta/}{http://www.atnf.csiro.au/research/pulsar/ppta/}
for details; and also the webpage of the International Pulsar Timing Array
\href{http://www.ipta4gw.org/}{http://www.ipta4gw.org/})

The PTA technique allows searchers of GW in frequencies ranging
from $10^{-9}-10^{-8}$ Hz. The sources of GWs being SMBHs
and stochastic waves of cosmological origin.

Finally, we consider very briefly the detection of GWs using the Cosmic Microwave Background (CMB).
GWs of very large wavelengths (which correspond to frequencies of $10^{-18}-10^{-15}$ Hz)
can in principle polarize the CMB. Therefore, a way to detect such GWs is to search for the
electromagnetic B-mode polarization \cite[see, e.g.,][]{2011RMxAC..40..299A}.
We refer the reader to the Planck Satellite
webpage (\href{http://planck.esa.int/}{http://planck.esa.int/}), for example,
for further details.

\section{Gravitational wave astronomy: what is already possible to do?}
\label{GWASTRO}

Even though we have not yet a direct detection of GWs, after
around 50 years of experimental research, it is already  possible
to do gravitational wave astronomy.

There is a very active community working in different
fronts of the data analysis research, namely, searches
for compact binaries, continuous waves, stochastic waves, and
burts. Some of these research activities incorporate also
electromagnetic counterparts (multimessage astronomy).

For the lack of space, it is considered here only some results
that come from the LIGO and VIRGO collaboration.
% Notice that so far, LIGO has six and VIRGO 3 science runs.

In a recent  paper by \cite{2012PhRvD..85h2002A}, it is presented the
search for GWs from low mass compact binary coalescence in
LIGO's sixth science run and Virgo's science runs 2 and 3 (S6-VSR2-3).

In table 1 it is presented the estimated coalescence rates
for NS-NS, NS-BH, BH-BH binary systems
\cite[see, e.g.,][and references therein for details]{2012PhRvD..85h2002A}.

\begin{table}[!ht] \label{coal}
\centering
\begin{tabular}{l|c c c}
Source & Low  & Realistic & High\\
\hline
\hline
NS-NS  & 0.01             & 1                & 10   \\
NS-BH  & $5\times 10^{-4}$ & 0.03             & 1    \\
BH-BB  & $10^{-4}$        & $5\times 10^{-3}$ & 0.03 \\
\hline
\hline
\end{tabular}
\caption{Estimated coalescence rates for NS-NS, NS-BH, BH-BH binary systems \cite[see, e.g.,][and references therein for details]{2012PhRvD..85h2002A} in $Mpc^{-3}\, Myr^{-1}$. With ``Realistic" meaning a realistic estimate, and ``High" and ``Low"
meaning the plausible limits of the range.}
\end{table}

As can been seen in \cite{2012PhRvD..85h2002A}, the S6-VSR2-3 constrains the coalescence
rates to $\sim 10^{-6}-10^{-5}-10^{-4}\; Mpc^{-3}\, yr^{-1}$ for BH-BH, NS-BH and NS-NS
binary systems, respectively. This is still far from the estimated coalescence rates.

Recall that the aLIGO will be able to see a volume of the universe around a
thousand larger than the initial LIGO, and its coalescence detection rates for compact binary
systems are estimated to be of tens per year \cite[see, e.g.,][]{2013PrPNP..68....1R}.
Therefore, the coalescing compact binaries will be very probably seen.

We refer the reader to the following articles for other interesting studies
concerning GW searches, namely:
a) ``searches for GWs from known pulsars with S5 LIGO data" \citep{2010ApJ...713..671A};
b) ``Directional Limits on Persistent Gravitational Waves Using LIGO S5 Science Data"
\citep{2011PhRvL.107A1102A};
c) ``All-sky search for gravitational-wave bursts in the second joint LIGO-Virgo run
Bursts" \citep {2012PhRvD..85l2007A}

An interesting study that makes uses of the electromagnetic radiation and the putative GW
counterpart (multimessage astronomy) refers to the GRB 070201, a short $\gamma$-ray burst. This GRB has a
reconstructed position consistent with Andromeda (M31). The absence of any signal in LIGO
data by the time of the burst and the proximity of M31 make it very unlikely that this
GRB was a binary merging in this galaxy \citep{2008ApJ...681.1419A}.

\section{Final remarks}
\label{Final}

It is (almost) a consensus that the era of GW astronomy has begun.
As we have seen, there is a very active search for GWs. Although,
there has been so far no detection, it is already
possible to constrain some parameters and models.

Also, we have seen that the most promising sources of GWs involve
compact objects; in particular the compact binary systems (NS-NS, BH-BH, BH-NS) are
the best bets for the first detection.

The so called ``multimessage astronomy" \cite[see, e.g.,][]{2011arXiv1105.5843C}, that incorporates
gravitational radiation with other messengers,
such as electromagnetic (ranging from radio to $\gamma$-rays)
and neutrino emission, is already a reality.

The simultaneous observation of electromagnetic waves, and also neutrino emissions, will be 
certainly important for the first direct detection of GWs. We refer the reader to
\href{http://toros.phys.utb.edu/TOROS}{http://toros.phys.utb.edu/TOROS} to see an example of
an astronomical observatory for follow-up of GW events.

Also, multimessenger observation could well help to answer some long standing questions, for example, what is the source of GRBs.

It is worth stressing that it is widely accepted that
in a few years, before the end of the present decade, the GWs
will be directly detected by the advanced configurations of
LIGO and VIRGO, for example. Recall that aLIGO will be able to see a volume a thousand larger than the initial LIGO.

To conclude, we are about to witness one of the most extraordinary discoveries of all times.

\agradecimientos
I would like to thank Dr. Gustavo E. Romero and the organizing
committees for inviting me to the plenary talk at AAA2012, Dr. Ricardo Morras for the arrangements concerning my trip to Mar del Plata, and Dr. Adrian Rovero for patiently wait for my contribution. Last, but not least, I would like to thank all of you and Dra. Paula Benaglia for the warm hospitality, and also many students and researchers for the interesting discussions. I am partially supported by FAPESP and CNPq.

\bibliographystyle{baaa}
\bibliography{deAraujo-Jose}
\end{document}